\documentclass{aa}  
\usepackage{graphicx}
\usepackage{txfonts}
\begin{document}
\title{The radio-loud active nucleus in the ``dark lens'' galaxy J1218+2953}

\author{S. Frey\inst{1,2}
       \and
       Z. Paragi\inst{3,2}
       \and
       R.M. Campbell\inst{3}
       \and 
       A. Mo\'or\inst{4}
       }

\institute{F\"OMI Satellite Geodetic Observatory, P.O. Box 585, 
          H-1592 Budapest, Hungary\\ 
          \email{frey@sgo.fomi.hu}
          \and
          MTA Research Group for Physical Geodesy and Geodynamics, P.O. Box 91, 
          H-1521 Budapest, Hungary
          \and
          Joint Institute for VLBI in Europe, Postbus 2, 
          7990 AA Dwingeloo, The Netherlands\\
          \email{zparagi@jive.nl, campbell@jive.nl}
          \and
          MTA Konkoly Observatory, P.O. Box 67, H-1525 Budapest, Hungary\\
          \email{moor@konkoly.hu}
          }

   \date{Received Dec 14, 2009; accepted Jan 22, 2010}

 
  \abstract
   {There is a possibility that the optically unidentified radio source J1218+2953 may act as a
gravitational lens, producing an optical arc $\sim$$4\arcsec$ away from the
radio position. Until now, the nature of the lensing object has been
uncertain since it is not detected in any waveband other than the radio. The
estimated high mass-to-light ratio could even allow the total mass of this
galaxy to be primarily in the form of dark matter. In this case, J1218+2953 could
be the first known example of a ``dark lens''.}
   {We investigate the nature of J1218+2953 by means of high-resolution radio imaging observations to determine whether there is a radio-loud active galactic nucleus (AGN) in the position of the lensing object.}
   {We report on Very Long Baseline Interferometry (VLBI) observations with the European VLBI Network (EVN) at 1.6 and 5~GHz.}
   {Our images, having
angular resolutions of $\sim$1 to $\sim$10 milli-arcseconds (mas), reveal a
rich and complex radio structure extending to almost $1\arcsec$. Based on its
radio spectrum and structure, J1218+2953 can be classified as a compact
steep-spectrum (CSS) source, and as a medium-size symmetric object (MSO). The
object harbours an AGN. It is also found as an X-ray source in the XMM-Newton EPIC (European Photon Imaging Cameras) instrument serendipitous source catalogue.}
   {Rather than being a dark lens, J1218+2953 is most likely a
massive, heavily obscured galaxy in which the nuclear activity is currently in
an early evolutionary stage.}


   \maketitle
%

\section{Introduction}
\label{intro}

It is now widely accepted that $\sim$95\% of the mass of galaxies and clusters is composed of some unknown form of dark matter. The observational evidence comes from the flat galactic rotation curves, the mass required for gravitational lensing, and the presence of hot X-ray-emitting gas in galaxy clusters (see e.g. Freese \cite{Free09} for a recent review). It is possible that empty dark matter halos may also exist. Strong gravitational lensing may in principle be a useful tool to detect massive dark matter halos that do not contain significant amounts of ordinary matter (Rusin \cite{Rusi02}). Hawkins (\cite{Hawk97}), based on a sample of supposedly lensed double quasars without visible lensing galaxies, envisaged a large number of dark galaxies, three times more than galaxies with normal mass-to-light ratios. However, those quasars later proved to be physical binaries rather than lensed images. In fact, the extensive systematic Cosmic Lens All-Sky Survey (CLASS) that identified gravitationally lensed objects in the radio, did not find any convincing evidence for dark lens among 22 lens systems (Jackson et al. \cite{Jack98}; Browne et al. \cite{Brow03}). 

Another suspected ``dark galaxy'', VIRGOHI~21 has earlier been
found in the Virgo Cluster, based on the broadening of its 21-cm H{\sc I} emission line
by the presumed rotation of the gravitationally bound neutral
hydrogen gas (Davies et al. \cite{Davi04}; Minchin et al. \cite{Minc05}). In this model, the total mass of the object
could reach $\sim$$10^{11}$~$M_{\odot}$, with a baryonic fraction at least ten
times smaller than observed in visible disk galaxies in the Universe
(Minchin et al. \cite{Minc07}). On both observational and theoretical grounds, others claim
that VIRGOHI~21 is in fact tidal debris left over from an interaction that
involved the nearby bright spiral galaxy NGC~4254 (Haynes et al. \cite{Hayn07}; Duc \& Bournaud \cite{Duc08}). The
existence of dark galaxies is therefore far from being established.

Recently Ryan et al. (\cite{Ryan08}) investigated a hypothesis that the optically unidentified radio source FIRST J121839.7+295325 (J1218+2953 hereafter) is strongly lensing a background galaxy. The arc-like lensed image is $\sim$$4\arcsec$ south-west of the radio source. Its estimated photometric redshift is $z\approx2.5$. However, the suspected foreground object is so far detected only in the radio. Its integrated flux density at $\nu=1.4$~GHz frequency is $S=33.9$~mJy in the Faint Images of the Radio Sky at Twenty-centimeters\footnote{\tt {http://sundog.stsci.edu}} (FIRST) survey (White et al. \cite{Whit97}). The source is unresolved in FIRST, with a deconvolved size smaller than $1\arcsec$. Other total radio flux density measure­ments found in the literature for this object at low frequencies are: 284~mJy at 74~MHz (Cohen et al. \cite{Cohe04}), 151~mJy at 151~MHz (Hales et al. \cite{Hale07}), and 129~mJy at 330~MHz (Westerbork Northern Sky Survey Catalogue, WENSS\footnote{\tt {http://cdsarc.u-strasbg.fr/viz-bin/Cat?VIII/62}}; de Bruyn et al. \cite{Bruy98}). A power-law fit to the total flux density data at these four frequencies gives a radio spectral index $\alpha=-0.7$ ($S\propto\nu^{\alpha}$).

Deep optical and infrared imaging (Russell et al. \cite{Russ08}; Ryan et al. \cite{Ryan08}) did not reveal any
counterpart to the radio source. The limiting magnitudes are 25.5, 22.0 and
20.7 in the $V$, $J$ and $H$ bands, respectively. Based on the 1.4-GHz radio
flux density and the optical non-detection (Russell et al. \cite{Russ08}), Ryan et al. (\cite{Ryan08}) place
the source in an approximate redshift range of $0.8 < z_{\rm{radio}} < 1.5$.
The lower limit derives from the optical non-detection, so if the object is an unusually
obscured galaxy, it could possibly be even closer.
The best-fit lens model assuming an isothermal ellipsoid
(Ryan et al. \cite{Ryan08}) provides a lensing galaxy with an Einstein radius of $1\farcs3$
and a dynamical mass $10^{12.5\pm0.5}$~$M_{\odot}$. The apparently very
large mass-to-light ratio led Ryan et al. (\cite{Ryan08}) to raise the possibility that
J1218+2953 may be in fact a galaxy dominated by dark matter.  Suspected
dark-matter galaxies, of which this could be a very rare (or the only) example,
would of course be impossible to image directly in the optical. 
Alternatively, the lensing radio source J1218+2953 may be a massive obscured galaxy with a central active galactic nucleus (AGN) visible in the radio. In this case, according to the optical non-detection and the lens model, the stellar mass in this galaxy would at best be only about 1\% of its total dynamical mass (Ryan et al. \cite{Ryan08}).

Firm evidence for a radio-loud AGN can be obtained with high-resolution radio
interferometric observations. Here we report on our dual-frequency Very Long
Baseline Interferometry (VLBI) experiments with the European VLBI Network (EVN), which have revealed a complex
structure on angular scales from $\sim$1 to several hundred
milli-arcseconds (mas) within J1218+2953. Our VLBI observations also provide an accurate
astrometric position of the source. We give the details of the experiments and the data
analysis, as well as our search for additional data at other wavebands in Sect.~\ref{observ}. We describe the observed radio structure of the
source in Sect.~\ref{structure} and discuss the possible implications of our
results in Sect.~\ref{discussion}. 

\section{Observations}
\label{observ}

\subsection{Radio interferometric imaging}

We observed J1218+2953 with the EVN at 1.6~GHz and
5~GHz frequencies, using the e-VLBI technique (Szomoru \cite{Szom08}).  Unlike
traditional VLBI, in which the remote radio telescopes of the
network record their signals onto physical media, which are shipped
to the central processor for subsequent correlation,
e-VLBI streams the signals over optical fibre networks directly to the central processor for
real-time correlation. This process vastly compresses the time-scale between
observations and the availability of the correlated data for
further analysis.
On 2009 January 23 we conducted an exploratory 1.6-GHz e-VLBI experiment.
This lasted only for 2
hours, but allowed us to verify that the source is sufficiently compact for VLBI detection.
Its radio emission was clearly detectable at $\sim$25~mas angular resolution,
and the resulting image revealed two major components separated by
$\sim$500~mas.  Based on these exploratory observations, we conducted
deeper dual-frequency observations.
To make a more sensitive 1.6-GHz VLBI image, we used
nine antennas of the EVN during an 8-hour observation on 2009 April 21.
Participating stations included Effelsberg (Germany),
Jodrell Bank Lovell Telescope, Cambridge, Darnhall (UK), Medicina (Italy),
Onsala (Sweden), Toru\'n (Poland), Arecibo (Puerto Rico) and the phased array
of the 14-element Westerbork Synthesis Radio Telescope (WSRT, the Netherlands).
At most of the antennas, a data transmission rate of 512~Mbit~s$^{-1}$ was
achieved in real time, which resulted in a total bandwidth of 64~MHz in both
left and right circular polarizations, using 2-bit sampling. 
We also carried out an 8-hour e-VLBI observation with EVN at 5~GHz on
2009 March 24. The maximum data rate was 1024~Mbit~s$^{-1}$, corresponding to
128~MHz bandwidth per polarization. The successfully participating radio
telescopes were Effelsberg, Jodrell Bank Mk2, Knockin (UK), Medicina, Onsala, 
Toru\'n, and the WSRT.
The correlation of the VLBI data from both observations took
place at the EVN MkIV Data Processor at the Joint Institute for VLBI in Europe
(JIVE) in Dwingeloo, the Netherlands.

We also analysed the synthesis array data recorded at the WSRT during the e-VLBI observations. Here, the source J1218+2953 appeared unresolved on arcsecond scales at both 1.6 and 5~GHz, allowing us to obtain simultaneous total flux density measurements. 

Because our target radio source is relatively weak, we observed in
phase-reference mode. This method is usually applied to increase the total
coherent integration time spent on the source and consequently to improve the
sensitivity of the observations. Phase-referencing is done by regularly
changing between the target source and a bright, compact reference
source lying nearby on the plane of the sky (e.g. Beasley \& Conway \cite{Beas95}). 
We chose J1217+3007, a BL Lac object with compact VLBI structure
(Bondi et al. \cite{Bond04}) having an angular separation of $0\fdg28$ from J1218+2953, as
the phase-reference calibrator.  We took the
position of J1217+3007 from
the US National Radio Astronomy Observatory (NRAO) Very Long Baseline Array
(VLBA) Calibrator
Survey\footnote{\tt {http://www.vlba.nrao.edu/astro/calib/index.shtml}},
which quoted right ascension and declination uncertainties of 0.21~mas and
0.41~mas, respectively.
The delay, delay rate and phase solutions
derived for the phase-reference calibrator were interpolated and applied to
J1218+2953 within the target--reference cycle time of $\sim$5 minutes. The
target source was observed for 3.5-minute intervals in each cycle. The total
observing time on J1218+2953 was nearly 5~h and 4~h at 1.6 GHz and 5 GHz,
respectively.

\begin{figure}
\centering
  \includegraphics[bb=67 170 523 625,width=85mm,angle=270,clip= ]{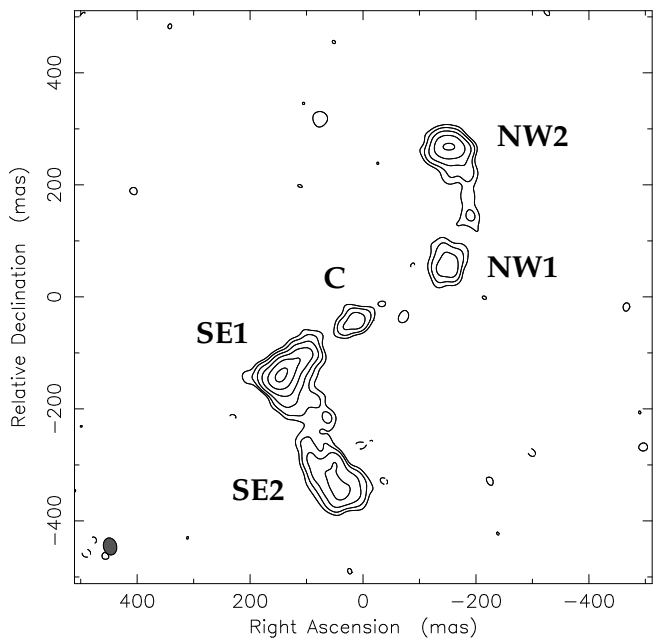}
  \caption{
The naturally weighted 1.6-GHz VLBI image of J1218+2953. The positive contour levels increase by a factor of 2. A Gaussian taper with the value of 0.5 at the projected baseline length of 10 million wavelengths (M$\lambda$) was applied to reduce the relative weight of the longest baselines (i.e., mainly the baselines from the European antennas to Arecibo). This provided better imaging of the extended emission. The lowest contours are drawn at $\pm70\;\mu$Jy/beam (negative contours are dashed). The peak
brightness is 2.67~mJy/beam. The Gaussian restoring beam is 30.9~mas~$\times$~23.3~mas at a major-axis position angle of $13{\degr}$.
   }
  \label{target-1.6GHz-tapered}
\end{figure}

\begin{figure}
\centering
  \includegraphics[bb=67 170 523 625,width=85mm,angle=270,clip= ]{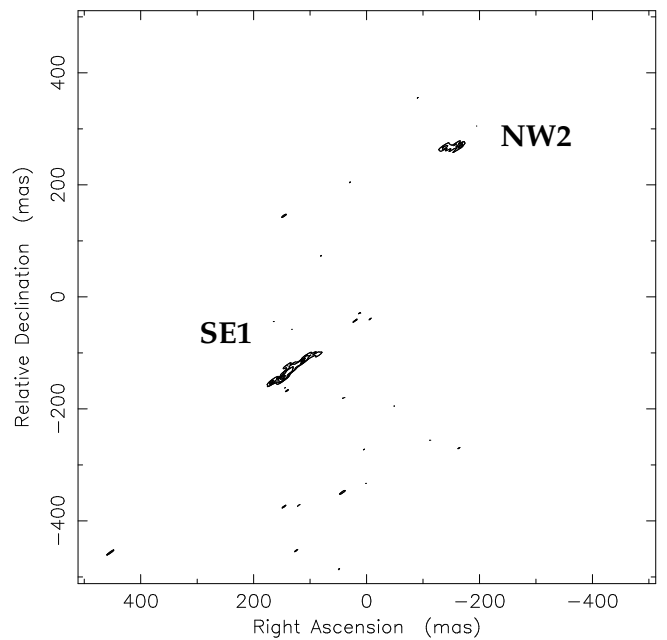}
  \caption{
The uniformly weighted 1.6-GHz VLBI image of J1218+2953. The positive contour levels increase by a factor of 2. The first contours are drawn at $-60$ and 60~$\mu$Jy/beam. The peak brightness is 0.62~mJy/beam. The Gaussian restoring beam is 15~mas~$\times$~2.2~mas at PA=$-54{\degr}$.
   }
  \label{target-1.6GHz-hires}
\end{figure}

\begin{figure}
\centering
  \includegraphics[bb=67 170 523 625,width=85mm,angle=270,clip= ]{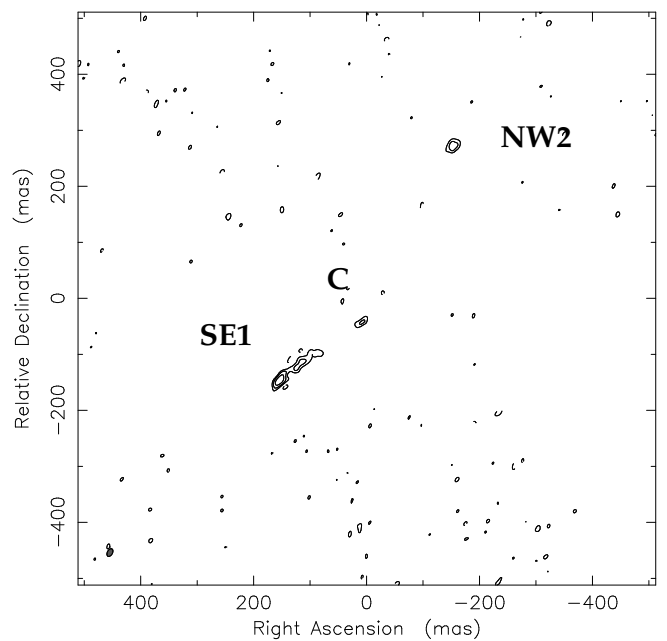}
  \caption{
The naturally weighted 5-GHz VLBI image of J1218+2953. The positive contour levels increase by a factor of 2. The first contours are drawn at $-50$ and 50~$\mu$Jy/beam. The peak brightness is 0.34~mJy/beam. The Gaussian restoring beam is 14.9~mas~$\times$~9.2~mas at PA=$-29{\degr}$.
   }
  \label{target-5GHz}
\end{figure}

We used the NRAO Astronomical Image Processing System (AIPS; e.g. Diamond \cite{Diam95}) for the data calibration. The visibility amplitudes were calibrated using system temperatures and antenna gains measured at the antennas. Fringe-fitting was performed for the calibrator (J1217+3007) and fringe-finder sources (J0927+3902, J1159+2914, J1407+2827) using 3-min solution intervals.
We exported the data to the Caltech Difmap package (Shepherd et al. \cite{Shep94}) for imaging. The conventional hybrid mapping procedure involving several iterations of CLEANing and phase (then amplitude) self-calibration resulted in the images and brightness distribution models for the calibrators. Overall antenna gain correction factors ($\sim$10\% or less) were determined and applied to the visibility amplitudes in AIPS.
We then repeated fringe-fitting for the phase-reference calibrator in AIPS, now
taking its CLEAN component model into account in order to compensate for
residual phase corrections resulting from its non-pointlike structure. The
solutions obtained were interpolated and applied to the target source data. The calibrated and phase-referenced
visibility data of J1218+2953, unaveraged in time, were also exported to Difmap
for imaging. 
The total intensity
images at 1.6~GHz (Fig.~\ref{target-1.6GHz-tapered}-\ref{target-1.6GHz-hires})
and 5~GHz (Fig.~\ref{target-5GHz}) were made after several cycles of CLEANing
in Difmap. The lowest contours are drawn at $\sim$$3\sigma$ image noise levels.
The coordinates in all images are relative to the {\it a priori} position taken from the FIRST catalogue ($\alpha_{\rm{0}} = 12^{\rm{h}}18^{\rm{m}}39\fs718$, $\delta_{\rm{0}} = +29{\degr}53{\arcmin}25\farcs58$). This position was used as the correlation phase center for both e-EVN experiments.

\subsection{Infrared and X-ray data}

At the position of J1218+2953, the archive of the Submillimetre Common-User Bolometer Array (SCUBA) at the James Clerk Maxwell Telescope (JCMT) does not contain data. To look for archival observations at shorter wavelengths, we searched the Spitzer Space Telescope archive for possible measurements at our target position. This resulted in two Infrared Array Camera (IRAC) images at 3.6 and 5.8~$\mu$m, which belong to the same observation (AOR key 12451840) targeting the nearby Seyfert galaxy Markarian~766 (NGC~4253). We started the data processing of these images with the BCD files (Basic Calibrated Data) and then performed additional corrections as described by Hora et al. (\cite{Hora08}). Since J1218+2953 was not detected in the final images, we derived upper limits at the position of our radio source of 0.030~mJy ($17\fm4$) and 0.28~mJy ($14\fm4$) at 3.6 and 5.8~$\mu$m, respectively. However, these limits are not very deep. 

The object 2XMM J121839.6+295327 found in the XMM-Newton EPIC (European Photon Imaging Cameras) instrument serendipitous source catalogue (Watson et al. \cite{Wats09}) is $2\farcs08$ away from the center of the radio source, to the north. This difference is comparable to the formal X-ray positional error ($1\farcs99$). Therefore we identify our radio source with the X-ray object. It is detected in the three highest-energy bands out of the five bands, with flux values of $(11.3\pm4.2) \times 10^{-19}$~W~m$^{-2}$ ($1-2$~keV band), $(17.3\pm9.7) \times 10^{-19}$~W~m$^{-2}$ ($2-4.5$~keV band), and $(222.3\pm82.5) \times 10^{-19}$~W~m$^{-2}$ ($4.5-12$~keV band). The total-band ($0.2-12$~keV) X-ray flux is $(258.8\pm83.3) \times 10^{-19}$~W~m$^{-2}$. 
Although the uncertainties are large, we can estimate the absorbing neutral hydrogen column density based on the detections in the 3 hardest EPIC bands, assuming a typical power-law spectral slope of 1.8. The value of $N_{\rm H}$$\sim$$10^{22}$~cm$^{-2}$ is about two orders of magnitude higher than the Galactic one.

\section{The radio structure of J1218+2953}
\label{structure}

Our tapered 1.6-GHz VLBI image (Fig.~\ref{target-1.6GHz-tapered}) reveals a rich and complex structure in an ``inverted S'' shape, spanning almost $0\farcs7$. This corresponds to a projected linear size of 5--6~kpc, if the source is indeed in the $0.8 < z_{\rm{radio}} < 1.5$ redshift range. (We assume a flat cosmological model with $H_{\rm{0}}=70$~km~s$^{-1}$~Mpc$^{-1}$, $\Omega_{\rm m}=0.3$ and $\Omega_{\Lambda}=0.7$.) The major components of the brightness distribution are labeled in the images. The two brightest ones (SE1 and NW2) seen also in the highest-resolution image (Fig.~\ref{target-1.6GHz-hires}) are on opposite sides of the approximate center of symmetry (component C). The 1.6-GHz data measured on the longest baselines to Arecibo indicate that there isn't any compact unresolved component visible with the sensitivity of those baselines (Fig.~\ref{target-1.6GHz-hires}). Therefore SE1 itself is in fact not a ``core--jet'' structure: the linear $\sim$100-mas long jet-like feature in the south-east ends in a ``hot spot''. Here the shape of the radio structure abruptly changes at right angles, ending up in component SE2. On the northwestern side of C, a similar but somewhat smoother directional change occurs between NW1 and NW2.

The sum of flux densities in the CLEAN components is 20.8~mJy. Our WSRT data
give 27~mJy total flux density, closer to the FIRST value (33.9~mJy). The
correlated flux density measured on only the shortest e-VLBI baseline in our
experiment (Jodrell Bank Lovell Telescope--Darnhall, 17.6~km) is also
consistent with this higher value. Therefore we believe that there is still
some extended emission in the $\sim$$1\arcsec$ angular scale which our e-VLBI  
observations resolved out.

The 5-GHz image (Fig.~\ref{target-5GHz}) shows only those components that have somewhat flatter spectra and compact structures not resolved out at this higher observing frequency. The sum of CLEANed flux densities is 1.6~mJy, while the WSRT data give 9~mJy total flux density. The difference is attributed to extended radio emission. The total flux density is consistent with the steep overall radio spectrum of the source.

The astrometric position of component C ($\alpha_{\rm{C}} = 12^{\rm{h}}18^{\rm{m}}39\fs7186$, $\delta_{\rm{C}} = +29{\degr}53{\arcmin}25\farcs537$) was determined with the AIPS task MAXFIT using the 5-GHz image, and is accurate to 1~mas. We suggest that this coincides with the center of the galaxy.

\section{Discussion}
\label{discussion}

Here we follow the assumption of Ryan et al. (\cite{Ryan08}) that the optical arc seen $\sim$$4\arcsec$ away from the
radio position of J1218+2953 is a result of gravitational lensing. However, we cannot provide any further proof for this 
scenario based on our own observations. We stress nevertheless that our interpretation of the compact radio source itself is independent of the lensing hypothesis.

\subsection{The nature of the radio source: lensed or lens?}

Are J1218+2953 and the optical arc all gravitationally lensed images of the
same background object, or does the compact radio source reside in the lens
forming the optical arc?  

A static gravitational lens would be achromatic and
conserve surface brightness in its images; these characteristics form the
basis for the principal objection to the former interpretation.
The radio flux density of the
images should scale with their size, and since the arc is elongated, its radio
flux density should be much higher than we observe in the compact
J1218+2953 components. However, the optical arc does not show sufficient
radio emission. At 1.6~GHz, both our VLBI radio source and the optical arc are located within the same WSRT restoring beam. Yet the measured WSRT flux density is only $\sim$25\% greater. 
In principle it might be possible that we see a gravitational lens system 
in which one of the ray paths suffers high obscuration in the optical, while the other one does in the radio, for example due to free-free absorption (e.g. Mittal et al. \cite{Mitt07}). However, we find this configuration very unlikely, because of the extended area of the arc that would require free-free absorption in order to mask a radio detection.

The structure seen in Fig.~\ref{target-1.6GHz-tapered} may suggest the 
appearance of two close images of a single bent structure in a background
source.  While the configuration
of an extended arc plus two more compact images is quite feasible 
(e.g. B0128+437; Biggs et al. \cite{Biggs04}, Biggs \cite{Biggs04evn}), 
conservation of surface brightness would imply that
the relative flux densities
of the two components in each ``image" should be similar.  
However, it is clear that NW1 is rather fainter than NW2, and the opposite
case holds for SE1 and SE2.  

It thus seems very unlikely that the radio structure of
J1218+2953 and the optical arc arise from graviational lensing of the same background source.
Therefore we interpret the host galaxy of J1218+2953 as (part of) the gravitational
lens forming the optical arc.

\begin{figure}
\centering
  \includegraphics[width=90mm,angle=0,clip= ]{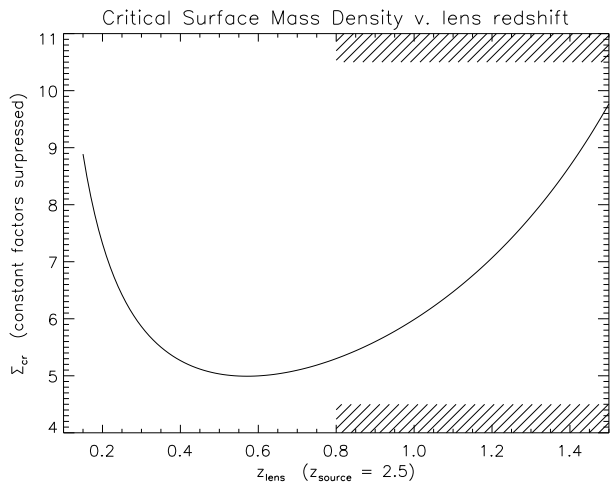}
  \caption{The critical surface mass density as a function of the lens
redshift for the case of a background source at $z=2.5$.
The constant factor ($cH_0/4\pi G$) in $\Sigma_{\rm cr}$ is supressed.
The hatched region along the abscissa is the range of $z_{\rm lens}$ 
considered by Ryan et al. (\cite{Ryan08}).
           }
  \label{crit.surf.mass.dens}
\end{figure}

Assuming a background object at $z=2.5$, even if the lensing galaxy is much closer than estimated by Ryan et al. (\cite{Ryan08}), i.e.
at $0.15<z<0.8$, the total lensing mass required to produce the same image-plane geometry would not change dramatically, only within a factor
of $\sim$2. 
Fig.~\ref{crit.surf.mass.dens} plots the critical surface mass density,
\begin{equation}
\frac{c^2}{4\pi G}\;\frac{D_{\rm s}}{D_{\rm d}D_{\rm ds}}, 
\end{equation}
supressing the explicit constants and the net factor of $H_0/c$ arising from 
the ratio of the usual lensing angular diameter distances.  
Therefore the interpretation of the VLBI radio source as belonging to
the lensing object is not terribly sensitive to the redshift of the lens,
even if the lensing galaxy moves significantly closer, reflecting higher 
obscuration in the optical.

Note that the center of the lens model (Fig.~2 of Ryan et al. \cite{Ryan08}) 
is offset from our radio position by $\sim$$2\farcs2$, beyond the $1\sigma$ positional uncertainty 
of the HST-WFPC2 observations ($\sim$$1\arcsec$; Russell et al. \cite{Russ08}).
In their model, Ryan et al. (\cite{Ryan08}) treated the position of the lensing galaxy, an isothermal ellipsoid, as a free parameter. 
The accurate coordinates derived from our VLBI observations could provide further constraints to refine the lens modeling, 
leading to somewhat different values for other parameters. On the other hand, similar radio sources are often associated with merging or disturbed-morphology hosts. Thus it is also possible that the VLBI structure does not coincide with 
the mass center of the galaxy.

\subsection{The classification of the radio source}

Based on its radio spectrum and structure, J1218+2953 can be classified as a compact steep-spectrum (CSS; e.g. O'Dea \cite{ODea98}) source, and a medium-size symmetric object (MSO; Fanti et al. \cite{Fant95}). These objects are believed to be young radio sources, recently (about $10^5$--$10^6$~years ago) triggered AGNs. Their host galaxies are the most easily identified in the nearby Universe. Beyond redshift $\sim$1, the hosts have typical $R$-band magnitudes around 24. These galaxies often have very red colors ($R-K>5$). (See e.g. de Vries \cite{Vrie03} and Holt \cite{Holt09} for reviews.)

The radio source J1218+2953 is probably entirely contained within its host
galaxy. A plausible assumption is that the mass center of the
host coincides with component C. The interaction of the outflowing plasma in the
jets with the dense interstellar medium could result in the observed two-sided
bent radio structure. 
This qualitative picture is consistent with the supposed
presence of a large mass of gas and dust in this ``dark'' galaxy. Notably, the
position angle of the practically linear inner part of the radio source
(SE1--NW1) is coincident with the minor axis of the gravitational lens model
(Fig.~2 of Ryan et al. \cite{Ryan08}). This is to be expected if we assume that the inner
jets mark the spin axis of the central black hole, which itself
coincides with the rotation axis of the entire galaxy.

\subsection{Is J1218+2953 a highly obscured galaxy?}

According to our VLBI imaging results, J1218+2953 can most naturally be identified with the lensing galaxy, if the optical arc is indeed due to gravitational lensing as we assume.
The lens is therefore not completely dark matter, but rather not (yet) detected in the optical and near-infrared wavebands. 
According to Ryan et al. (\cite{Ryan08}), this galaxy has an extremely high (over $\sim100$) dynamical-to-stellar mass ratio. However,
due to the suspected high obscuration, 
the stellar mass could take more of the total mass in this galaxy.
It is feasible to expect mid- and far-infrared emission from the dust, and (hard) X-rays from the high-energy photons originating from the accretion disk around the central black hole that can penetrate the dense material in the galaxy.

The early evolutionary stages of the growth of massive black holes are highly obscured. The model of Fabian (\cite{Fabi99}) accounts for the observed hard X-ray background and states that the growth of the central black hole continues until the obscuring gas is ejected from the galaxy. Before that, the absorbed radiation from the obscured AGN is re-emitted mostly in the far-infrared and sub-millimeter bands. Direct X-ray radiation from around the accreting black hole may be observed above 30~keV. According to theoretical models, CSS sources should generally be strong X-ray emitters. However, at $z\approx1$ and beyond, the angular resolution of current X-ray telescopes is not sufficient to resolve the expected kpc-scale structure spatially (Siemiginowska \cite{Siem09}, and references therein). 

The flux density upper limit in the 3.6-$\mu$m Spitzer band and the measured XMM-Newton EPIC flux allow us to compare our source with a sample of obscured AGNs selected by Eckart et al. (\cite{Ecka10}) from moderately deep fields observed by both the Spitzer Space Telescope and the Chandra X-ray Observatory. While our source fits in the general X-ray--infrared dependence, our upper limit indicates that J1218+2953 could be relatively faint in the mid-infrared compared to other obscured AGNs.

\begin{figure}
\centering
  \includegraphics[width=70mm,angle=270,clip= ]{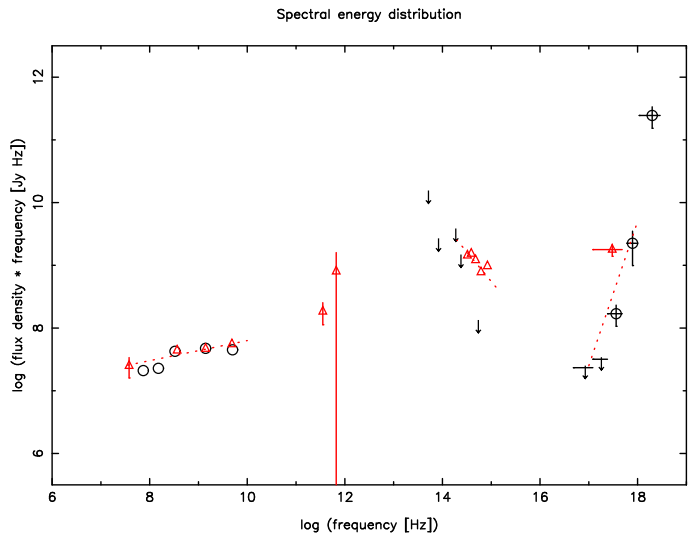}
  \caption{
The SED of J1218+2953. Measured values are indicated with open circles, upper limits with arrows. The values for RX~J1011.2+5545 (Barcons et al. \cite{Barc98}) divided by 5 are plotted with triangles. The data are taken form the NASA/IPAC Extragalactic Database (NED). Dashed lines characterise the radio (left) and optical (middle) part of the comparison SED. The dashed line at the X-rays (right) is the approximation of the spectrum of RX~J1011.2+5545 from Fig.~6 of Barcons et al. (\cite{Barc98}).}
  \label{sed}
\end{figure}

We constructed the spectral energy distribution (SED) of J1218+2953 from the measured data in the radio and X-rays, and from the upper limits available in the optical and near-infrared (Fig.~\ref{sed}). For comparison, we plotted the SED of an X-ray-selected moderately obscured radio-loud AGN at $z=1.25$, RX~J1011.2+5545 (Barcons et al. \cite{Barc98}). The flux densities of the comparison object were scaled down by a factor of 5 throughout the whole electromagnetic spectrum, such that the 1.4-GHz flux densities of the two sources were aligned. While the two resulting SEDs are remarkably similar in the radio and X-ray bands, J1218+2953 appears at least an order of magnitude fainter in the optical, suggesting an additional extinction $\Delta A_{\rm V}>2\fm5$.

Without dust obscuration, Ryan et al. (\cite{Ryan08}) place the dynamical mass-to-light ratio in the range of $\sim$30 to 150 $M_{\odot} L_{\odot}^{-1}$ for J1218+2953. A strong extinction of $\sim$$3^{\rm m}$ would imply higher luminosity and consequently a mass-to-light ratio consistent with the typical values for other galaxies.
The $A_{\rm V}/N_{\rm H}$ ratio in our case ($\sim$$10^{-22}$~mag~cm$^2$) is less than the standard Galactic value of $5.3 \times 10^{-22}$~mag~cm$^2$, but there is observational evidence that this ratio is nearly always lower in other AGNs, typically by factors of $\sim$3 to 100 (Maiolino et al. \cite{Maio01}). Thus for J1218+2953, there is a rough agreement between the likely amount of extinction and the column density of the absorbing neutral hydrogen inferred from the X-ray data.

\section{Conclusion}
\label{conclusion}

Our EVN observations have revealed that the optically unidentified radio source J1218+2953 has a radio-loud AGN in its center. The tapered 1.6-GHz e-VLBI image (Fig.~\ref{target-1.6GHz-tapered}) shows a complex, two-sided radio structure. The angular size of the radio source is less than $1\arcsec$. Although the redshift of the object is unknown, we used the range $0.8 < z_{\rm{radio}} < 1.5$ estimated by Ryan et al. (\cite{Ryan08}) to conclude that the source is confined to a sub-galactic projected linear size, 5--6~kpc. Our higher-resolution images at 1.6~GHz (Fig.~\ref{target-1.6GHz-hires}) and 5~GHz (Fig.~\ref{target-5GHz}) indicate two ``hot spots'' and a possible weak central component. The latter may mark the location of the massive black hole in the center of this galaxy. We determined the position of this radio component to an accuracy of 1~mas.

The suspected lensing object thus coincides with a compact radio source associated with an AGN. According to well-established models, its radio jet activity is driven by accretion onto a supermassive (typically $\sim$$10^{8}$~$M_{\odot}$) black hole (see e.g. Urry \& Padovani \cite{Urry95} for a review). Moreover, we found that the object also emits X-rays. It is therefore detected in two different electromagnetic wavebands, which indicates that its total mass could not be exclusively in the form of dark matter. 

Based on its radio spectrum and morphology, J1218+2953 can be classified as a compact steep-spectrum (CSS) source, and as a medium-size symmetric object (MSO). Our object is unique in a sense that the CSS samples studied to date contain 
radio sources at least twice as bright. The faintest CSS sample of Tschager et al. (\cite{Tsch03})
was compiled from unresolved WENSS sources with at least 250~mJy flux density
at 330~MHz. However, even those sources usually lack high-resolution VLBI data,
therefore their pc-scale properties are unknown. 

A few relatively weak MSOs have been observed by Kunert-Bajraszewska et al. (\cite{Kune05}) with the NRAO Very Large Array (VLA) and
the UK Multi-element Radio Linked Interferometer Network (MERLIN), with an eventual goal of establishing a more complete evolutionary scheme for radio sources.
In our case, J1218+2953 has attracted much attention as an potential ``dark'' gravitational lens, not
because it was a faint CSS. 
The complex structure we observe draws attention to the importance of in-depth high-resolution
studies of faint CSS sources, and obscured radio-loud X-ray-selected AGNs. These would help provide a better understanding of the
early evolution of radio AGNs, the interaction between the host galaxies and
the expanding radio sources, and AGN feedback processes. 

Our results render the ``dark lens'' interpretation of Ryan et al. (\cite{Ryan08}) unlikely. Supported by the X-ray detection of J1218+2953, we suspect that it is a heavily obscured galaxy in which the nuclear activity is in an early evolutionary stage. Deeper observations in the future in the mid-infrared, far-infrared, and sub-millimeter bands might be able to detect the source and put more constraints on its spectral energy distribution. 

\begin{acknowledgements}
We are grateful to the chair of the EVN Program Committee, Tiziana Venturi, for granting us short exploratory e-VLBI observing time in January 2009.  
The EVN is a joint facility of European, Chinese, South African and other radio astronomy institutes funded by their national research councils. 
The Westerbork Synthesis Radio Telescope is operated by the ASTRON (Netherlands Institute for Radio Astronomy) with support from the Netherlands Foundation for Scientific Research (NWO).
This effort is supported by the European Community Framework Programme 7, Advanced Radio Astronomy in Europe, grant agreement no.\ 227290, and the Hungarian Scientific Research Fund (OTKA, grant no.\ K72515). The e-VLBI developments in the EVN are supported by the EC DG-INFSO funded Communication Network Developments project `EXPReS', Contract No.\ 02662 ({\tt http://www.expres-eu.org/}). This research has made use of the NASA/IPAC Extragalactic Database (NED) which is operated by the Jet Propulsion Laboratory, California Institute of Technology, under contract with the National Aeronautics and Space Administration.
\end{acknowledgements}

\end{document}